# Blockchained Federated Learning for Threat Defense

**Konstantinos Demertzis**[1,2]

[1] Laboratory of Complex Systems, Department of Physics, Faculty of Sciences, International Hellenic University, Kavala Campus, St. Loukas, 65404, Greece; kdemertzis@teiemt.gr

[2] Faculty of Mathematics Programming and General Courses, Department of Civil Engineering, School of Engineering, Democritus University of Thrace, Kimmeria, Xanthi, Greece, kdemertz@fmenr.duth.gr

**Abstract:** Given the increasing complexity of threats in smart cities, the changing environment and the weakness of traditional security systems, which in most cases fail to detect serious threats such as zero-day attacks, the need for alternative more active and more effective security methods keeps increasing. Such approaches are the adoption of intelligent solutions to prevent, detect and deal with threats or anomalies under the conditions and the operating parameters of the infrastructure in question. Intelligent systems are capable, of displaying logical, empirical, and non-human decision-making, since they are trained appropriately by historical data representative of the problem they are trying to solve. In most cases, it is either not possible or it is inappropriate to centrally store all smart cities data. Thus, we should perform real-time knowledge mining and we should obtain a subset of a data flow containing a small but recent percentage of observations. This fact raises serious objections to the accuracy and reliability of the employed intelligent system classifiers, who have been tame over time and they become incapable of detecting serious threats. This research paper introduces the development of an intelligent Threat Defense system, employing Blockchain Federated Learning, which seeks to fully upgrade the way passive intelligent systems operate, aiming at implementing an Advanced Adaptive Cooperative Learning (AACL) mechanism for smart cities networks. The AACL is based on the most advanced methods of computational intelligence, while ensuring privacy and anonymity for participants and stakeholders. The proposed framework combines Federated Learning for the distributed and continuously validated learning of the tracing algorithms. Learning is achieved through encrypted smart contracts within the blockchain technology, for unambiguous validation and control of the process. The aim of the proposed Framework is to intelligently classify smart cities networks traffic derived from Industrial IoT (IIoT) by Deep Content Inspection (DCI) methods, in order to identify anomalies that are usually due to Advanced Persistent Threat (APT) attacks.

## 1. Introduction

A smart city is an urban development vision to integrate multiple information and communication technology solutions in a secure fashion to manage a city's assets – the city's assets include, but not limited to, local departments information systems, schools, libraries, transportation systems, hospitals, power plants, water supply networks, waste management, law enforcement, and other community services [1]. The goal of building a smart city is to improve quality of life by using technology to improve the efficiency of services and meet residents' needs [2], [3]. Information and communication technology allows city officials to interact directly with the community and the city infrastructure and to monitor what is happening in the city, how the city is evolving, and how to enable a better quality of life [4].

Through the use of sensors integrated with real-time monitoring systems, data are collected from citizens and devices - then processed and analyzed. The information and knowledge gathered are keys to tackling inefficiency. Also information and communication technology is used to enhance quality, performance and interactivity of urban services, to reduce costs and resource consumption and to improve contact between citizens and government. Smart city applications are developed with the goal of improving the management of urban flows and allowing for real time responses to challenges.

In a smart city, energy, water, transportation, public health [5] and safety, and other key services are managed in concert to support smooth operation of critical infrastructure while providing for a clean, economic and safe environment in which to live, work and play. Timely logistics information will be gathered and supplied to the public by all means available, but particularly through social media networks. Conservation, efficiency and safety will all be greatly enhanced. The energy infrastructure is arguably the single most important feature in any city. If unavailable for a significant enough period of time, all other functions will eventually cease [6].

A smart grid alone does three things [7]–[9]. First, it modernizes power systems through self-healing designs, automation, remote monitoring and control, and establishment of microgrids. Second, it informs and educates consumers about their energy usage, costs and alternative options, to enable them to make decisions autonomously about how and when to use electricity and fuels. Third, it provides safe, secure and reliable integration of distributed and renewable energy resources. All these add up to an energy infrastructure that is more reliable, more sustainable and more resilient. Thus, a smart grid sits at the heart of the smart city, which cannot fully exist without it.

Smart cities depend on a smart grid to ensure resilient delivery of energy to supply their many functions, present opportunities for conservation, improve efficiencies and, most importantly, enable coordination between urban officialdom, infrastructure operators, those responsible for public safety and the public. The smart city is all about how the city "organism" works together as an integrated whole and survives when put under extreme conditions. Energy, water, transportation, public health and safety, and other aspects of a smart city are managed in concert to support smooth operation of critical infrastructure while providing for a clean, economic and safe environment in which to live, work and play [10].

In smart cities, information security plays a major role in protecting the higher levels of confidentiality, availability, and integrity as well as the stability that national services and organizations need to support sustainable and livable smart environments. The biggest security challenges for the smart city environments, which can be summarized as the following [1], [11], [12]:

1. *Large and Complex attack surface*: The smarter the cities, the more systems and "systems of systems" they will incorporate, increasing the risk and impact of an attack, thus requiring better control and visibility. Furthermore, what adds to the smart city systems complexity is the integration between vendor's solutions, especially during fast evolving technological transformations.
2. *Insufficient oversight and organization*: Complex systems will then require stronger management and governance capabilities; in addition, keeping leadership fully knowledgeable of complex occurrences requires more resources and capabilities.

At the core of smart city operation depends on control systems called Supervisory Control and Data Acquisition (SCADA) that monitor and control the physical infrastructure. SCADA systems that tie together decentralized facilities such as power, oil, gas pipelines, water distribution and wastewater collection systems were designed to be open, robust, and easily operated and repaired, but not necessarily secure [9], [13], [14]. The move from proprietary technologies to more standardized and open solutions together with the increased number of connections between SCADA systems, office networks and the Internet has made them more vulnerable to types of network attacks that are relatively common in computer security.

In particular, security researchers are concerned about [13], [15]:
1. the lack of concern about security and authentication in the design, deployment and operation of some existing SCADA networks,
2. the belief that SCADA systems have the benefit of security through obscurity through the use of specialized protocols and proprietary interfaces,
3. the belief that SCADA networks are secure because they are physically secured,
4. the belief that SCADA networks are secure because they are disconnected from the Internet.

The security of SCADA systems is important because compromise or destruction of a smart city systems would impact multiple areas of society far removed from the original compromise. For example, a blackout caused by a compromised electrical SCADA system would cause financial losses to all the customers that received electricity from that source.

There are many threat vectors to a modern SCADA system. One is the threat of unauthorized access to the control software, whether it be human access or changes induced intentionally or accidentally by virus infections and other software threats residing on the control host machine. Another is the threat of packet access to the network segments hosting SCADA devices.

In many cases, the control protocol lacks any form of cryptographic security, allowing an attacker to control a SCADA device by sending commands over a network. In many cases SCADA users have assumed that having a VPN offered sufficient protection, unaware that security can be trivially bypassed with physical access to SCADA-related network jacks and switches. Also, there are destructive cyber-attacks against SCADA systems as Advanced Persistent Threats (APT), were able to take over the PLCs controlling the centrifuges, reprogramming them in order to speed up the centrifuges, leading to the destruction of many and yet displaying a normal operating speed in order to trick the centrifuge operators and finally can not only shut things down but can alter their function and permanently damage industrial equipment.

Industrial control vendors suggest approaching SCADA security like Information Security with a defense in depth strategy that leverages common IT practices. The reliable function of SCADA systems in our modern infrastructure may be crucial to public health and safety [13], [16], [17].

**2. Decentralized Ecosystem**

Industry 4.0, commonly referred to as the fourth industrial revolution, is concerned with the trend of automation and data exchange in industrial ecosystem including the smart cities automations [12], [18]. It includes technologies like artificial intelligence, cyber-physical systems, IoT/IIoT, cloud and cognitive computing.

The cyber-physical systems located inside modular structured smart factories monitor and supervise physical processes, create a virtual copy of the physical world, and take decentralized decisions. Through the IIoT, the cyber-physical systems communicate and collaborate in real time with each other and with people both internally and through organizational services which are offered and used by participants in the production chain. This vision enables the manufacturing sector to make tremendous breakthroughs, gain significant extroversion, and develop activities that were previously impossible.

Cyber criminals can have access to the IIoT process with serious perhaps incalculable consequences, most stakeholders are demanding high-performance security solutions, to be able to cope with the dangers and to shield their infrastructures [19].

Production facilities, industrial systems and smart city networks in general need a different kind of protection from company networks, as conventional security solutions, such as virus scanners or conventional firewalls, do not meet industry standards and requirements.

The network systems that control the process and operation of smart cities have continuous access to the internet, to the IIoT that belong to them, and to the information and data of the company or organization they belong to. Such access, digital communication and connectivity, improve the efficiency of their operation but at the same time, they pose significant challenges of safeguarding these infrastructures in terms of their digital identity and integrity [11].

Internet interconnection and data exchange increase the risk of attacks, which may be aimed at stealing, manipulating and spying data or how to manipulate them. This can lead to the loss of sensitive company data, the sabotage of individual machines, or even the cessation of a whole production line. One very important fact that exacerbates the situation is that machines and devices in modern industrial facilities are not designed to be securely connected, making them particularly vulnerable to cybernetics. The growing number of such attacks in the production facilities confirms this fact.

Smart city networks should be able to achieve its goals and it is particularly important to ensure procedures and to resolve cyber-security issues, in order to ensure the operational continuity and productivity of the systems related to that environment.

IIoT security event monitoring and digital threat detection systems receive huge amounts of data per time unit from heterogeneous specialized equipment systems that are interconnected.

Because in most cases it is not possible or inappropriate to centrally store all historical data, it is imperative to extract real-time knowledge over data flows that contain a small but recent percentage of observations of the overall set. In order to exploit these flows, and to achieve timely behavioral prediction and optimal decision-making in dynamic displacement and feedback environments, where the most up-to-date data is usually most important, it is necessary to process and analyze them using alternative, more active and more effective safety methods

Algorithms that are employed to resolve data flow problems, should be dynamically adapted to new standards or data, or when the data itself is generated as a function of time. Specifically, and on the basis that the available data are staggered in a successive series by calculating the error in each iteration, the aim of these algorithms is to minimize the cumulative error for all iterations [20], [21].

Intelligent real-time data flow analysis systems are able of displaying logical, empirical and human decision-making capabilities, when they are properly trained by representative historical datasets. Moreover, they are the most suitable to be used in industrial environments.

However, even these intelligent algorithms are challenged and controlled by a variety of possible factors that are mainly related to the reliability and accuracy of their categorization. Some of the most significant problems encountered during knowledge mining from data flows, are related to the high speed at which information arrive and to the natural trend of data to evolve over time, resulting in classifiers' lability because of the constant information change (concept drift) [22], [23].

The solution proposed herein, in order to resolve the above problems, is the development of an intelligent active security system, for the classification of network traffic derived from the IIoT. This can be achieved by using DCI methods and by performing anomaly detection. These potential anomalies are usually due to APTs attacks. The Federated Learning technique should be applied [24].

Development of a model by using the centralized training learning approach, requires the gathering of training data in a machine or a data center.

This central training approach is intrusive, since practically device users, should exchange their privacy or sensitive data by sending them to central entities for training.

On the other hand, federated learning is a decentralized training approach that allows devices located in different geographic locations to benefit from the acquisition and use of a well-and-ever, upgraded, trained learning model. In addition, this learning approach, allows all personal data and personal information that may be contained in the device to be retained, as they do not have to send them to a central entity to provide training data [25].

In particular, the Federated Learning method, allows devices to download and operate a current, trained machine learning model, located in a central entity.

They use it with the data available in their device, by improving learning from their data and then by sending it to the central entity, which summarizes the changes, creating a small focused update. The updated and improved version of the shared model, which is basically the average of user updates, is sent to users who constantly have access to a continuously upgraded model. There are various building blocks for which there should be trust and consensus among the parties involved in training of a machine learning model such as data acquisition, training, regularization and optimization [26].

For example, if an organization shares a set of data with a group of scientists, there is a silent relationship of trust between them. If it is violated, it can affect the result of the project. Similarly, in a more technical context, learning algorithms must undergo multiple training, regularization, and optimization cycles, in which teams focus on setting the model's hyperparameters. Basically, there is no clear way for the parties to collaborate in a sure and secure way, as there is no historical record or a complete record of the way they should act, so that users and data scientists cannot get information that is likely to concern them.

In the introduced model, we propose the use of smart contracts, within the function of blockchain, thus ensuring unambiguous validation and control of the processes among stakeholders [27], [28]. This model, seeks to fully upgrade the mode of operation of passive intelligent systems, aiming at the development of an advanced mechanism of adaptive collaborative learning. It offers fully personalized solutions and is based on the most advanced methods of computational intelligence, while ensuring privacy and anonymity for the participants [29].

This research paper seeks to understand the interactions between smart cities assets and cyber-attacks and proposes a novel computational intelligence cyber-threat defense system for smart city assets. The proposed approach builds a model that correlates a specialized smart city asset such as a smart energy grid and an APT cyber-attack. Using state-of-the-art methodologies analyzes big data from city networks, identify and defense these attack patterns.

More specifically, this paper proposes an intelligent Threat Defense system employing Blockchained Federated Learning, implementing an AACL mechanism that ensuring privacy and anonymity for participants and stakeholders on smart city networks. The proposed framework intelligently manage and classify IIoT traffic in order to identify anomalies and to defend against sophisticated cyber-attacks.

## 3. Proposed Framework

The proposed methodology is related to a Federated Learning architectural modeling. It aims to the development of a high quality and precision central model, where training data remains distributed over several IIoT devices, with possibly unreliable and relatively slow network connections. Correspondingly, the training of the model is realized by employing smart contracts within the blockchain technology. The model involves the development of an intelligent, multilevel industrial network analysis and protection mechanism, which allows the following:

1. The recognition of protocols and applications in DCI traffic.
2. The analysis of extracted data.
3. The depiction of anomalies in industrial IIoT devices.
4. Preemptive protection of IIoT from APTs attacks.

Respectively, it will provide real-time information about the state of the network and it will allow early detection of problems that may arise from infected machines, incorrect settings, or cyber-attacks.

In particular, the proposed architecture is based on four fundamental building blocks:

1. *Blockchain Server* – The consensus mechanism used by Blockchain Server:
   a. Every instance only has one authority validating transactions.
   b. Instead of one single central ledger, each authority controls their own instance. Instances can connect to each other.
   c. Different transactions will be validated by different authorities depending on the assets being exchanged.
   d. Every asset issuer has full control on the transactions relevant to that asset.
2. *Public_Key_Server* – to create public and private keys to ensure the integrity of various components of an intelligent model.
3. *Federated_Learning_Server* – it runs in blockchain and controls the execution of the different parts of an intelligent distributed learning application. The specific server will contain a suitable application that will communicate with the Public_Key_Server, in order to create keys and to deliver the respective ones. It also communicates with the Miners, selecting the upgrades and distributing the updated model.
4. *Miners* – It is an application hosted in individual devices. It detects possible upgrades of the intelligent model and it communicates regularly with the Federated Learning_Server, by sending the local upgraded model and by receiving the corresponding generic model.

5. *Machine_Leraning_Server* – It contains the machine learning algorithms that can be trained from user data. It also includes the original model which is downloaded by the users.

These five servers allow the implementation of an innovative communication channel in which data scientists, security and networks' engineers can collaborate in implementing cooperative learning security applications. Moreover, this architecture ensures the privacy and integrity of data and existing models without having to rely on a centralized training learning architecture.

**4. Application Scenario**

An *IIoT_Sender*, wishes to interact by sending a datastream with s*tream_ ID* 1029 in the *IIoT_ Receiver*. The Data Streams receive a whereas while the specific action of the shipment is monitored by the IIoT rule on whether that particular transaction can take place.

The IIoT rule activates the *Smart_Contract ID* (9009). The particular Smart Contract ID (9009), routes Stream IDs (6006) to a cloud service, where features extraction is performed, and then the trained Machine Learning classifier checks whether the DataStream is normal or abnormal. This specification results defined in the training process for the characterization of the network's traffic and it is measured in RMSE. If the traffic is considered normal, then there is communication with the *IIoT_Receiver*, otherwise the communication is hidden and an alert is sent to further inspection. The following figure presents the communication process between the devices, based on the DCI Contract.

The proposed scenario in pseudocode format is described in the following algorithm 1:

Algorithm 1. The proposed approach

*#MachineAccount*
IIoT_thing;
*#MachineAddress*
FE80:0000:0000:0000:0202:B3FF:FE1E:8329;
*#MachineInternals*
Energy (GeV) = 1.320 | Lifetime (hours) = 9658.150 | Current (mA) = 4.210
Last Refill = 23 May 22:11 | Mode = standby | Feedback Status = On
*#MachineStatus*
Publisher; Sender;
*#DataStream*
Size = 125 kb; TimeStamp = 201802305221100;
*#DataStreamID*
ID = 1029;
*# FeaturesOfTrafficFlow*
command_address;
response_address;
command_memory;
response_memory;
command_memory_count;

```
response_memory_count;
comm_read_function;
comm_write_fun;
resp_read_fun;
resp_write_fun;
sub_function;
command_length;
resp_length;
control_mode;
control_scheme;
pump;
crc_rate;
measurement;
#Functions
    function DataSender (MachineAccount, MachineAddress, MachineStatus);
    function DataReceiver (MachineAccount, MachineAddress, MachineStatus);
    function DataTransaction (DataStream, DataStreamID);
    function DataFlow (TrafficFlow, TrafficFlowID);
    function TransactionSession (Session, SessionID);
    function FeaturesOfTrafficFlow;
    function AnomalyDetection;
#Loop
    start
    if
        DataSender to DataReceiver a DataTransaction
    then DataFlow process to FeaturesOfTrafficFlow
    and FeaturesOfTrafficFlow to AnomalyDetection
    else if
    AnomalyDetection normal then TransactionSession
    end
```

Some applications that can take advantage of the proposed architecture are presented below.

1. *Machine to Machine Transactions*. Can be involved in organizing the production process by automatically interacting with the machines and exchanging messages about the products' readiness to move to the next stage of production.
2. *Machine Maintenance*. It will be able to perform autonomous scheduling of spare parts' requests, rebooting of equipment, shutdowns to replace problematic or obsolete hardware, software upgrades, and maintenance works' automation in general. All of the above will be easily and safely allowed.
3. *Traceability*. This architecture is also suitable for the development of traceability applications, related to industrial products and supply chain. Specifically, within an intelligent industrial environment, production logs can be kept between consumers and producers so that it is known, for example, which factory and in particular which machines at the plant were used to manufacture a particular product.

4. *Product Certification.* This can eliminate the need for physical certificates that may be prone to falsification.
5. *Predictive Manufacturing.* The proposed architecture ensures the integrity of provided services and the professional confidentiality of the parties involved.
6. f. *Reputation.* It can contribute significantly, in the management of reputation, related to a variety of performance parameters such as delivery times, customer reviews and supplier ratings.

## 5. Conclusions

This research paper presented an innovative blockchained federated learning for threat defense framework, based on sophisticated computational intelligence methods [30]. The most important innovation of the proposed system is the strengthening of the blockchain network by deploying federating learning, which does not behave as a supporting framework, but as an active structural component of the smart city networks. It is a threat defense framework, which programmatically implements the bi-directional agreement, based on federated learning. Another very important innovation is the employment of a machine learning implementation of an anomaly detection system, solving a multidimensional and complex security problem related to the IIoT ecosystem.

It would be important for the proposed framework to be expanded by employing methods of self-improvement and automatic redefinition of its parameters. In this way, the full automation of APT attacks detection, will become possible.

**Conflicts of Interest:** The authors declare no conflict of interest.

11 of 12[7]  J. Lin and L. Liu, "Research on Security Detection and Data Analysis for Industrial Internet," in *2019 IEEE 19th International Conference on Software Quality, Reliability and Security Companion (QRS-C)*, Jul. 2019, pp. 466–470, doi: 10.1109/QRS-C.2019.00089.

[8]  C. Zhou, Z. Wang, W. Huang, and Y. Guo, "Research on Network Security Attack Detection Algorithm in Smart Grid System," in *2017 International Conference on Computer Technology, Electronics and Communication (ICCTEC)*, Dec. 2017, pp. 1407–1410, doi: 10.1109/ICCTEC.2017.00307.

[9]  K. Demertzis, L. S. Iliadis, and V.-D. Anezakis, "An innovative soft computing system for smart energy grids cybersecurity," *Adv. Build. Energy Res.*, vol. 12, no. 1, pp. 3–24, Jan. 2018, doi: 10.1080/17512549.2017.1325401.

[10] S. Tousley and S. Rhee, "Smart and Secure Cities and Communities," in *2018 IEEE International Science of Smart City Operations and Platforms Engineering in Partnership with Global City Teams Challenge (SCOPE-GCTC)*, Apr. 2018, pp. 7–11, doi: 10.1109/SCOPE-GCTC.2018.00008.

[11] A. Banafa, "2 The Industrial Internet of Things (IIoT): Challenges, Requirements and Benefits," in *Secure and Smart Internet of Things (IoT): Using Blockchain and AI*, River Publishers, 2018, pp. 7–12.

[12] T. Gebremichael *et al.*, "Security and Privacy in the Industrial Internet of Things: Current Standards and Future Challenges," *IEEE Access*, vol. 8, pp. 152351–152366, 2020, doi: 10.1109/ACCESS.2020.3016937.

[13] G. Falco, C. Caldera, and H. Shrobe, "IIoT Cybersecurity Risk Modeling for SCADA Systems," *IEEE Internet Things J.*, vol. 5, no. 6, pp. 4486–4495, Dec. 2018, doi: 10.1109/JIOT.2018.2822842.

[14] S. Ghosh and S. Sampalli, "A Survey of Security in SCADA Networks: Current Issues and Future Challenges," *IEEE Access*, vol. 7, pp. 135812–135831, 2019, doi: 10.1109/ACCESS.2019.2926441.

[15] E. Irmak and I. Erkek, "An overview of cyber-attack vectors on SCADA systems," in *2018 6th International Symposium on Digital Forensic and Security (ISDFS)*, Antalya, Mar. 2018, pp. 1–5, doi: 10.1109/ISDFS.2018.8355379.

[16] R. Ankele, S. Marksteiner, K. Nahrgang, and H. Vallant, "Requirements and Recommendations for IoT/IIoT Models to automate Security Assurance through Threat Modelling, Security Analysis and Penetration Testing," in *Proceedings of the 14th International Conference on Availability, Reliability and Security*, New York, NY, USA, Aug. 2019, pp. 1–8, doi: 10.1145/3339252.3341482.

[17] H. Geng, "THE INDUSTRIAL INTERNET OF THINGS (IIoT)," in *Internet of Things and Data Analytics Handbook*, Wiley, 2017, pp. 41–81.

[18] K. R. Choo, S. Gritzalis, and J. H. Park, "Cryptographic Solutions for Industrial Internet-of-Things: Research Challenges and Opportunities," *IEEE Trans. Ind. Inform.*, vol. 14, no. 8, pp. 3567–3569, Aug. 2018, doi: 10.1109/TII.2018.2841049.

[19] K. Demertzis and L. Iliadis, "A Bio-Inspired Hybrid Artificial Intelligence Framework for Cyber Security," in *Computation, Cryptography, and Network Security*, N. J. Daras and M. Th. Rassias, Eds. Cham: Springer International Publishing, 2015, pp. 161–193.

[20] F. Mercaldo, F. Martinelli, and A. Santone, "Real-Time SCADA Attack Detection by Means of Formal Methods," in *2019 IEEE 28th International Conference on Enabling Technologies: Infrastructure for Collaborative Enterprises (WETICE)*, Jun. 2019, pp. 231–236, doi: 10.1109/WETICE.2019.00057.

[21] S. Raza, L. Wallgren, and T. Voigt, "SVELTE : Real-time Intrusion Detection in the Internet of Things," *Ad Hoc Netw.*, vol. 11, no. 8, pp. 2661–2674, 2013.

[22] K. Demertzis, L. Iliadis, and V. Anezakis, "MOLESTRA: A Multi-Task Learning Approach for Real-Time Big Data Analytics," in *2018 Innovations in Intelligent Systems and Applications (INISTA)*, Jul. 2018, pp. 1–8, doi: 10.1109/INISTA.2018.8466306.